\documentclass[aps,prl,twocolumn,groupedaddress]{revtex4}

\usepackage{array,hhline,dcolumn} 
\usepackage{rotating} 

\newcommand{\gtwid}{\mathrel{\raise.3ex\hbox{$>$\kern-.75em\lower1ex\hbox{$\sim$}}}}
\newcommand{\ltwid}{\mathrel{\raise.3ex\hbox{$<$\kern-.75em\lower1ex\hbox{$\sim$}}}}

\begin{document}
%


\title{Improved Search for $\bar \nu_\mu \rightarrow \bar \nu_e$ Oscillations in the MiniBooNE Experiment }

\author{
        A.~A. Aguilar-Arevalo$^{12}$, 
        B.~C.~Brown$^{6}$, L.~Bugel$^{11}$,
	G.~Cheng$^{5}$, E.~D.~Church$^{16}$, J.~M.~Conrad$^{11}$,
	R.~Dharmapalan$^{1}$, 
	Z.~Djurcic$^{2}$, D.~A.~Finley$^{6}$, R.~Ford$^{6}$,
        F.~G.~Garcia$^{6}$, G.~T.~Garvey$^{9}$, 
        J.~Grange$^{7}$,
        W.~Huelsnitz$^{9}$, C.~Ignarra$^{11}$, R.~Imlay$^{10}$,
        R.~A. ~Johnson$^{3}$, G.~Karagiorgi$^{5}$, T.~Katori$^{11}$,
        T.~Kobilarcik$^{6}$, 
        W.~C.~Louis$^{9}$, C.~Mariani$^{15}$, W.~Marsh$^{6}$,
        G.~B.~Mills$^{9}$,
	J.~Mirabal$^{9}$,
        C.~D.~Moore$^{6}$, J.~Mousseau$^{7}$, 
        P.~Nienaber$^{14}$, 
        B.~Osmanov$^{7}$, Z.~Pavlovic$^{9}$, D.~Perevalov$^{6}$,
        C.~C.~Polly$^{6}$, H.~Ray$^{7}$, B.~P.~Roe$^{13}$,
        A.~D.~Russell$^{6}$, 
	M.~H.~Shaevitz$^{5}$, 
        J.~Spitz$^{11}$, I.~Stancu$^{1}$, 
        R.~Tayloe$^{8}$, R.~G.~Van~de~Water$^{9}$, 
        D.~H.~White$^{9}$, D.~A.~Wickremasinghe$^{3}$, G.~P.~Zeller$^{6}$,
        E.~D.~Zimmerman$^{4}$ \\
\smallskip
(The MiniBooNE Collaboration)
\smallskip
}
\smallskip
\smallskip
\affiliation{
$^1$University of Alabama; Tuscaloosa, AL 35487 \\
$^2$Argonne National Laboratory; Argonne, IL 60439 \\
$^3$University of Cincinnati; Cincinnati, OH 45221\\
$^4$University of Colorado; Boulder, CO 80309 \\
$^5$Columbia University; New York, NY 10027 \\
$^6$Fermi National Accelerator Laboratory; Batavia, IL 60510 \\
$^7$University of Florida; Gainesville, FL 32611 \\
$^8$Indiana University; Bloomington, IN 47405 \\
$^9$Los Alamos National Laboratory; Los Alamos, NM 87545 \\
$^{10}$Louisiana State University; Baton Rouge, LA 70803 \\
$^{11}$Massachusetts Institute of Technology; Cambridge, MA 02139 \\
$^{12}$Instituto de Ciencias Nucleares, Universidad Nacional Aut\'onoma de M\'exico, D.F. 04510, M\'exico \\
$^{13}$University of Michigan; Ann Arbor, MI 48109 \\
$^{14}$Saint Mary's University of Minnesota; Winona, MN 55987 \\
$^{15}$Center for Neutrino Physics; Virginia Tech; Blacksburg, VA 24061\\
$^{16}$Yale University; New Haven, CT 06520\\
}

\date{\today}

\begin{abstract}
The MiniBooNE experiment at Fermilab reports results from an 
analysis of  $\bar \nu_e$ appearance data from $11.27 \times 10^{20}$ protons
on target in antineutrino mode, an increase of approximately a factor of two over
the previously reported results.  An event excess of $78.4 \pm 28.5$ events ($2.8 \sigma$)
is observed in the energy range $200<E_\nu^{QE}<1250$~MeV.  If interpreted in a
two-neutrino oscillation model, $\bar{\nu}_{\mu}\rightarrow\bar{\nu}_e$, the best oscillation
fit to the excess has a probability of 66\% while
the background-only fit has a $\chi^2$-probability of 0.5\% relative to the best
fit.  The data are consistent with antineutrino 
oscillations in the $0.01 < \Delta m^2 < 1.0$ eV$^2$ range and 
have some overlap 
with the
evidence for antineutrino oscillations from the Liquid Scintillator 
Neutrino Detector (LSND).  All of the major backgrounds are constrained by in-situ 
event measurements so non-oscillation explanations would need to invoke new anomalous
background processes.  The neutrino mode running also shows an excess
at low energy of $162.0 \pm 47.8$ events ($3.4 \sigma$) but the energy
distribution of the excess is marginally compatible with a simple two neutrino oscillation
formalism.  Expanded models with several sterile neutrinos can reduce 
the incompatibility by allowing for CP violating effects between neutrino 
and antineutrino oscillations.
\end{abstract}

\pacs{14.60.Lm, 14.60.Pq, 14.60.St}

\keywords{Suggested keywords}
\maketitle


There is growing evidence for short-baseline neutrino anomalies occurring at an $L/E_\nu \sim 1$ m/MeV,
where $E_\nu$ is the neutrino energy and $L$ is the distance that the neutrino travelled before detection.
These anomalies include the excess of events observed by the LSND \cite{lsnd} and MiniBooNE 
\cite{mb_osc,mb_lowe,mb_osc_anti}
experiments and the deficit of events observed by reactor \cite{reactor}
and radioactive-source experiments \cite{radioactive}.
There have been several attempts to interpret these anomalies in terms of 3+N 
neutrino oscillation models 
involving three active neutrinos and N additional sterile neutrinos 
\cite{sorel,karagiorgi,giunti,kopp,white_paper,3+2}. 
(Other explanations include, for example, Lorentz violation 
\cite{lorentz} and sterile neutrino decay \cite{sterile_decay}.)
A main goal of MiniBooNE was 
to confirm or refute 
the evidence for neutrino oscillations from LSND.
Of particular importance is the MiniBooNE search
for $\bar{\nu}_{\mu}\rightarrow\bar{\nu}_e$ oscillations
since this was the channel where LSND observed an
apparent signal.  
This paper presents improved results and an oscillation analysis
of the MiniBooNE $\bar \nu_e$ appearance data,
corresponding to $11.27 \times 10^{20}$ POT
in antineutrino mode, which is approximately twice the antineutrino
data reported previously \cite{mb_osc_anti}. 

Even though the first goal of this article is a presentation of the improved
antineutrino results, a secondary goal is to contrast and compare these
results with improved MiniBooNE neutrino measurements and, therefore, the details of 
both the neutrino and antineutrino analysis will be given.  Since the original neutrino
result publication \cite{mb_lowe}, improvements to the analysis have been made that
affect both the  $\nu_e$  and $\bar \nu_e$ appearance search.  These improvements
are described and used in the analyses presented here.

The neutrino (antineutrino) flux is produced by 8 GeV protons from the Fermilab Booster
interacting on a beryllium target inside a magnetic focusing horn set at positive 
(negative) polarity. In neutrino (antineutrino) mode, positively 
(negatively) charged mesons 
produced in p-Be interactions are focused in the forward direction 
and subsequently decay primarily into $\nu_\mu$ ($\bar{\nu}_{\mu}$). The flux of
neutrinos and antineutrinos of all flavors is simulated
using information from external measurements \cite{mb_flux}.
In neutrino mode, the $\nu_\mu$, $\bar \nu_\mu$, $\nu_e$, and $\bar \nu_e$ flux
contributions at the detector are 93.5\%, 5.9\%, 0.5\%, and 0.1\%, respectively.
In antineutrino mode, the $\bar \nu_\mu$, $\nu_\mu$, $\bar \nu_e$, and $\nu_e$ flux
contributions at the detector are 83.7\%, 15.7\%, 0.4\%, and 0.2\%, respectively.
The $\nu_\mu$ and $\bar{\nu}_{\mu}$ fluxes peak at approximately 600 MeV and 400 MeV, respectively. 

The MiniBooNE detector is described in detail in reference \cite{mb_detector}. 
The detector is located 541 m from the beryllium target and consists of
a 40-foot diameter sphere filled with 806 tons of pure mineral oil (CH$_{2}$). Neutrino interactions in the 
detector produce charged particles (electrons, muons, protons, pions, and kaons) 
which in turn produce scintillation and Cherenkov light 
detected by the 1520 8-inch photomultiplier tubes (PMTs) that line the interior of the detector and
an optically isolated outer veto region. Event reconstruction and particle identification are derived from
the hit PMT charge and time information.  In particular, the reconstructed neutrino energy, $E_\nu^{QE}$, 
uses the measured energy and angle of the outgoing muon 
or electron assuming charged-current quasi-elastic (CCQE) kinematics for the event.

The signature of $\nu_\mu \rightarrow \nu_e$ and
$\bar \nu_\mu \rightarrow \bar \nu_e$ oscillations 
is an excess of $\nu_e$ and $\bar \nu_e$-induced CCQE 
events. Reconstruction \cite{mb_recon} and selection requirements of these 
events are almost identical to those from previous 
analyses \cite{mb_lowe,mb_osc_anti} with an average reconstruction efficiency 
of $\sim 10-15\%$ for events generated over the entire volume of the
detector. Recent improvements to the analysis include a
better determination of the intrinsic $\nu_e$ background from $K^+$ decay 
through the measurement of high-energy neutrino events in the SciBooNE experiment \cite{sciboone_kaon},
a better determination of NC $\pi^0$ and external event backgrounds in antineutrino
mode due to the increase in statistics of the antineutrino mode data sample, and the use of a likelihood fit with
frequentist corrections from fake data studies for both the neutrino-mode and antineutrino-mode analyses.
The detector cannot distinguish
between neutrino and antineutrino interactions 
on an event-by-event basis. However, the fraction of CCQE events in antineutrino (neutrino) mode that
are due to wrong-sign neutrino (antineutrino) events was determined from the angular 
distributions of muons created in CCQE interactions and
by measuring CC single $\pi^+$ events \cite{wrong_sign}.

The predicted $\nu_e$ and $\bar{\nu}_e$ CCQE background events for 
the neutrino oscillation energy range $200<E_\nu^{QE}<1250$~MeV 
are shown in Table \ref{signal_bkgd} for both neutrino mode and antineutrino mode. 
MiniBooNE does not have the electron versus gamma particle identification capabilities to 
determine whether observed 
events are due to charged-current (CC) electron events, as
expected for an oscillation signal or intrinsic beam 
$\nu_e/\bar\nu_e$ background, or to background gamma events
from neutral-current (NC) interactions in the detector or interactions in the external 
surrounding material.  
The estimated size of the intrinsic $\nu_e$ and gamma backgrounds 
are tied to MiniBooNE event measurements and uncertainties due to these constraints 
are included in the analysis.  
The intrinsic $\nu_e/\bar\nu_e$ background from muon decay is
directly related to the large sample of 
observed $\nu_\mu / \bar\nu_\mu$ events since these events constrain the muons 
that decay in the 50 m decay region.  
(The $\nu_\mu / \bar\nu_\mu$ CCQE data sample,
in the $200<E_\nu^{QE}<1900$ MeV energy range,
includes 115,467 and 50,456 neutrino and antineutrino events, respectively .)  
This constraint is accomplished using a joint fit of
the observed $\nu_\mu\ / \bar\nu_\mu$ events and the $\nu_e/\bar\nu_e$ events assuming 
that there are no substantial $\nu_\mu\ / \bar\nu_\mu$ disappearance oscillations.  The other
intrinsic background $\nu_e$ component from K-decay is constrained by fits to kaon production
data and the recent SciBooNE measurements \cite{sciboone_kaon}.
Other backgrounds from mis-identified $\nu_{\mu}$ or $\bar{\nu}_{\mu}$ 
\cite{mb_numuccqe,mb_numuccpi} events are also constrained by the observed CCQE sample.
The gamma background from NC $\pi^0$ production mainly from $\Delta$ decay 
or $\Delta \rightarrow N\gamma$ radiative
decay \cite{hill_zhang} is constrained by the associated large two-gamma data sample (mainly from $\Delta$ 
production) observed in the MiniBooNE data \cite{mb_pi0}.  
In effect, an in-situ NC $\pi^0$ rate is measured and applied to the analysis.  Single-gamma backgrounds
from external neutrino interactions (``dirt" backgrounds) are estimated using topological and spatial cuts
to isolate these events whose vertex is near the edge of the detector and
point towards the detector center \cite{mb_lowe}.

\begin{table}[t]
\vspace{-0.1in}
\caption{\label{signal_bkgd} \em The expected (unconstrained) number of events
for the $200<E_\nu^{QE}<1250$~MeV neutrino oscillation 
energy range from all of the backgrounds in the $\nu_e$ and $\bar{\nu}_e$ 
appearance analysis and for an example 0.26\% oscillation probability averaged over neutrino energy
for both neutrino mode and antineutrino mode.  The table also shows the diagonal-element systematic uncertainties whose effects become substantially reduced in the oscillation fits when correlations between energy bins and between the electron and muon neutrino events are included. 
}
\small
\begin{ruledtabular}
\begin{tabular}{ccc}
Process&Neutrino Mode&Antineutrino Mode \\
\hline
$\nu_\mu$ \& $\bar \nu_\mu$ CCQE & 37.1 $\pm$ 9.7 & 12.9 $\pm$ 4.3 \\
NC $\pi^0$ & 252.3 $\pm$ 32.9 & 112.3 $\pm$ 11.5 \\
NC $\Delta \rightarrow N \gamma$ & 86.8  $\pm$12.1 & 34.7 $\pm$ 5.4 \\
External Events & 35.3 $\pm$ 5.5 & 15.3 $\pm$ 2.8 \\
Other $\nu_\mu$ \& $\bar \nu_\mu$ & 45.1 $\pm$ 11.5 & 22.3 $\pm$ 3.5 \\
\hline
$\nu_e$ \& $\bar \nu_e$ from $\mu^{\pm}$ Decay & 214.0 $\pm$ 50.4 & 91.4 $\pm$ 27.6 \\
$\nu_e$ \& $\bar \nu_e$ from $K^{\pm}$ Decay & 96.7  $\pm$ 21.1 & 51.2 $\pm$ 11.0 \\
$\nu_e$ \& $\bar \nu_e$ from $K^0_L$ Decay & 27.4 $\pm$ 10.3 & 51.4 $\pm$ 18.0 \\
Other $\nu_e$ \& $\bar \nu_e$ & 3.0 $\pm$ 1.6 & 6.7 $\pm$ 6.0 \\
\hline
Total Background &797.7&398.2 \\
\hline
0.26\% $\bar{\nu}_{\mu}\rightarrow\bar{\nu}_e$ & 233.0 & 100.0 \\
\end{tabular}
\vspace{-0.2in}
\end{ruledtabular}
\normalsize
\end{table}

Systematic uncertainties are determined by considering the predicted
effects on the $\nu_\mu$, $\bar{\nu}_{\mu}$, $\nu_e$, and $\bar{\nu}_e$ CCQE rate 
from variations of parameters.
These include uncertainties in the neutrino and antineutrino flux estimates, 
uncertainties in neutrino cross sections, most of which are determined by 
in-situ cross-section 
measurements at MiniBooNE \cite{mb_numuccqe,mb_pi0}, uncertainties due to nuclear effects, and uncertainties in 
detector modeling and reconstruction. 
A covariance matrix in bins of $E^{QE}_{\nu}$ is constructed 
by considering the variation from each source of systematic uncertainty on the $\nu_e$ and $\bar{\nu}_e$ CCQE signal, background, and 
$\nu_\mu$ and $\bar{\nu}_{\mu}$ CCQE prediction as a function of $E_{\nu}^{QE}$.
This matrix includes correlations between any of the $\nu_e$ and $\bar{\nu}_e$ CCQE signal and background and 
$\nu_\mu$ and $\bar{\nu}_{\mu}$ CCQE samples, and is used in the $\chi^2$ calculation of the oscillation fits.

\begin{figure}[tbp]
\vspace{+0.1in}
\centerline{\includegraphics[angle=0, width=7.0cm]{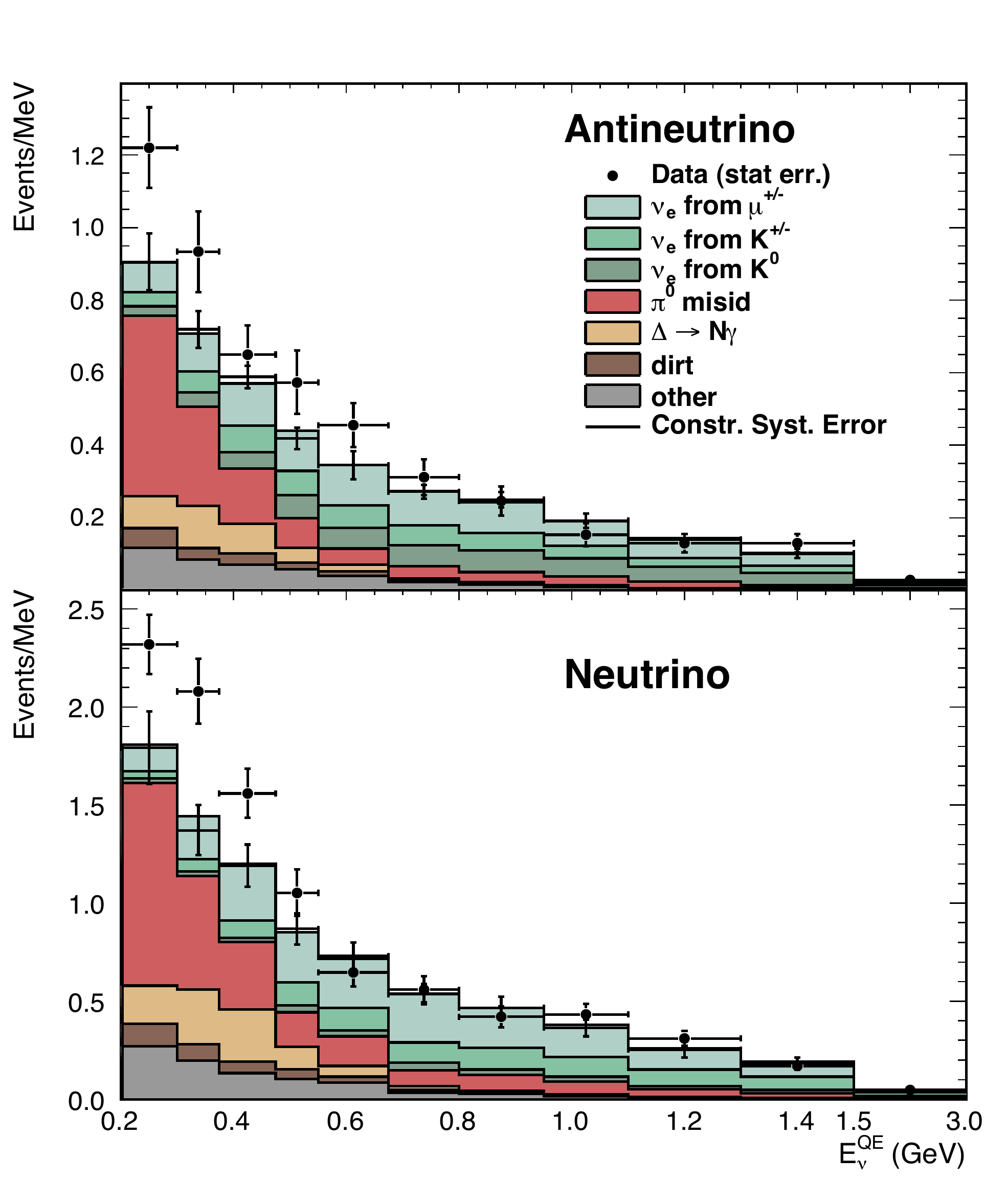}}
\vspace{-0.2in}
\caption{The antineutrino mode (top) and neutrino mode (bottom)  
$E_\nu^{QE}$ distributions 
for ${\nu}_e$ CCQE data (points with statistical errors) and background (histogram with systematic errors).} 
\label{excessnat}
\vspace{-0.2in}
\end{figure}

\begin{figure}[tbp]
\vspace{-0.0in}
\centerline{\includegraphics[angle=0, width=7.5cm]{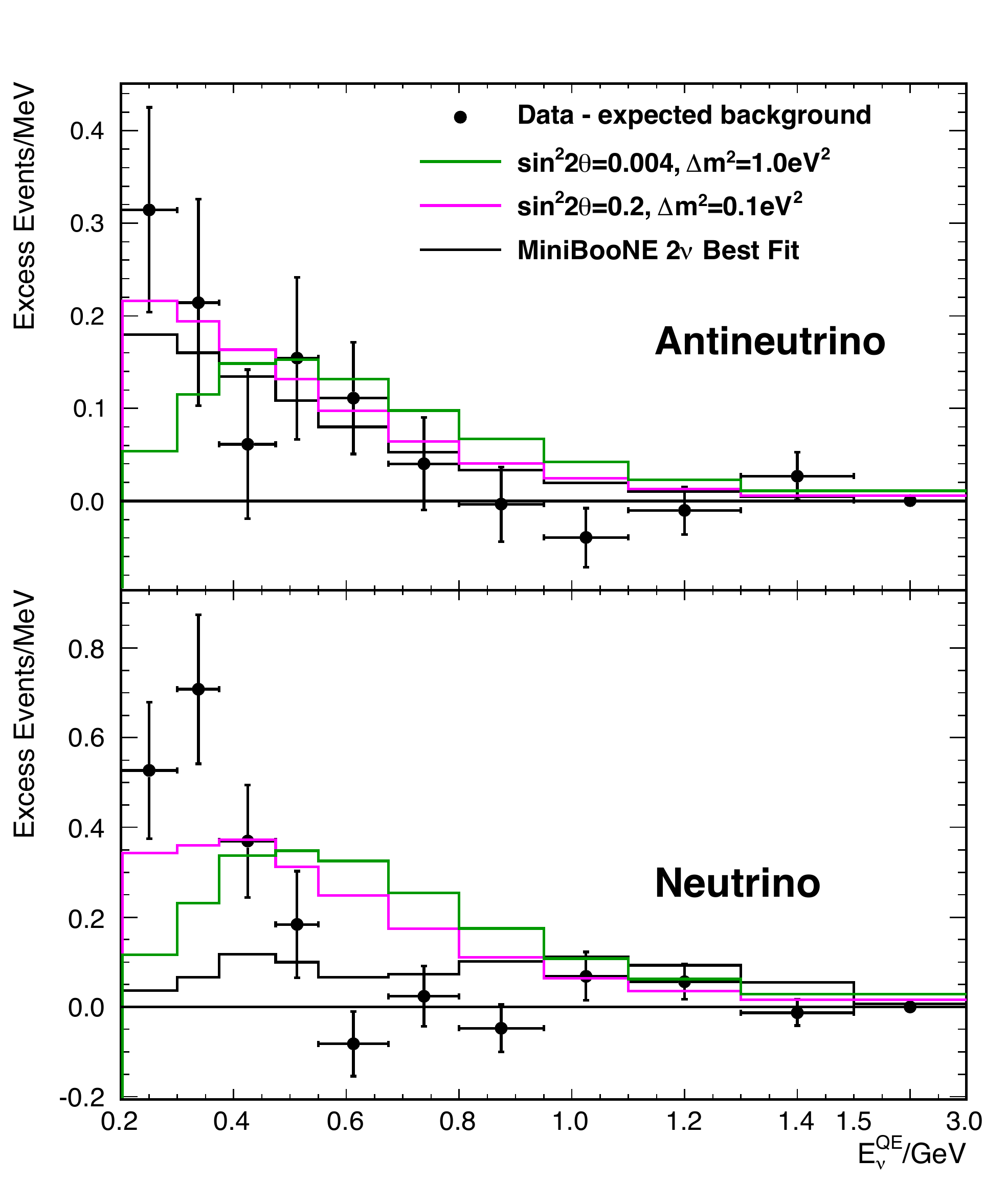}}
\vspace{-0.2in}
\caption{The antineutrino mode (top) and neutrino mode (bottom)
event excesses as a function of $E_\nu^{QE}$. (Error bars include
both the statistical and systematic uncertainties.) Also shown are the 
expectations from the best two-neutrino fit for each mode and 
for two example sets of oscillation parameters.}
\label{excessnab}
\vspace{-0.1in}
\end{figure}

Fig. \ref{excessnat} (top) shows the $E_\nu^{QE}$ distribution for 
$\bar{\nu}_e$ CCQE data and background in
antineutrino mode over the full available energy range. 
Each bin of reconstructed $E_\nu^{QE}$
corresponds to a distribution of ``true'' generated neutrino energies,
which can overlap adjacent bins.
In antineutrino mode, a total of 478 data events pass 
the $\bar\nu_e$ event selection requirements with $200<E_\nu^{QE}<1250$~MeV, 
compared to a background expectation of $399.6 \pm 20.0 (stat.) \pm 20.3 (syst.) $ events.
For assessing the probability that the expectation fluctuates up to this
478 observed value, the excess is then $78.4 \pm 28.5$ events
or a 2.8$ \sigma$ effect. 
Fig.~\ref{excessnab} (top) shows the event excess 
as a function of $E_\nu^{QE}$ in antineutrino mode.

Many checks have been performed on the data, including 
beam and detector stability checks that show that the neutrino event 
rates are stable to $<2\%$ and
that the detector energy response is stable to $<1\%$ over the entire run. 
In addition, the fractions of neutrino
and antineutrino events are stable over energy 
and time, and the inferred external event rate corrections are similar
in both neutrino and antineutrino modes. 

\begin{figure}[tbp]
\vspace{-0.25in}
 \centerline{\includegraphics[width=6.9cm]{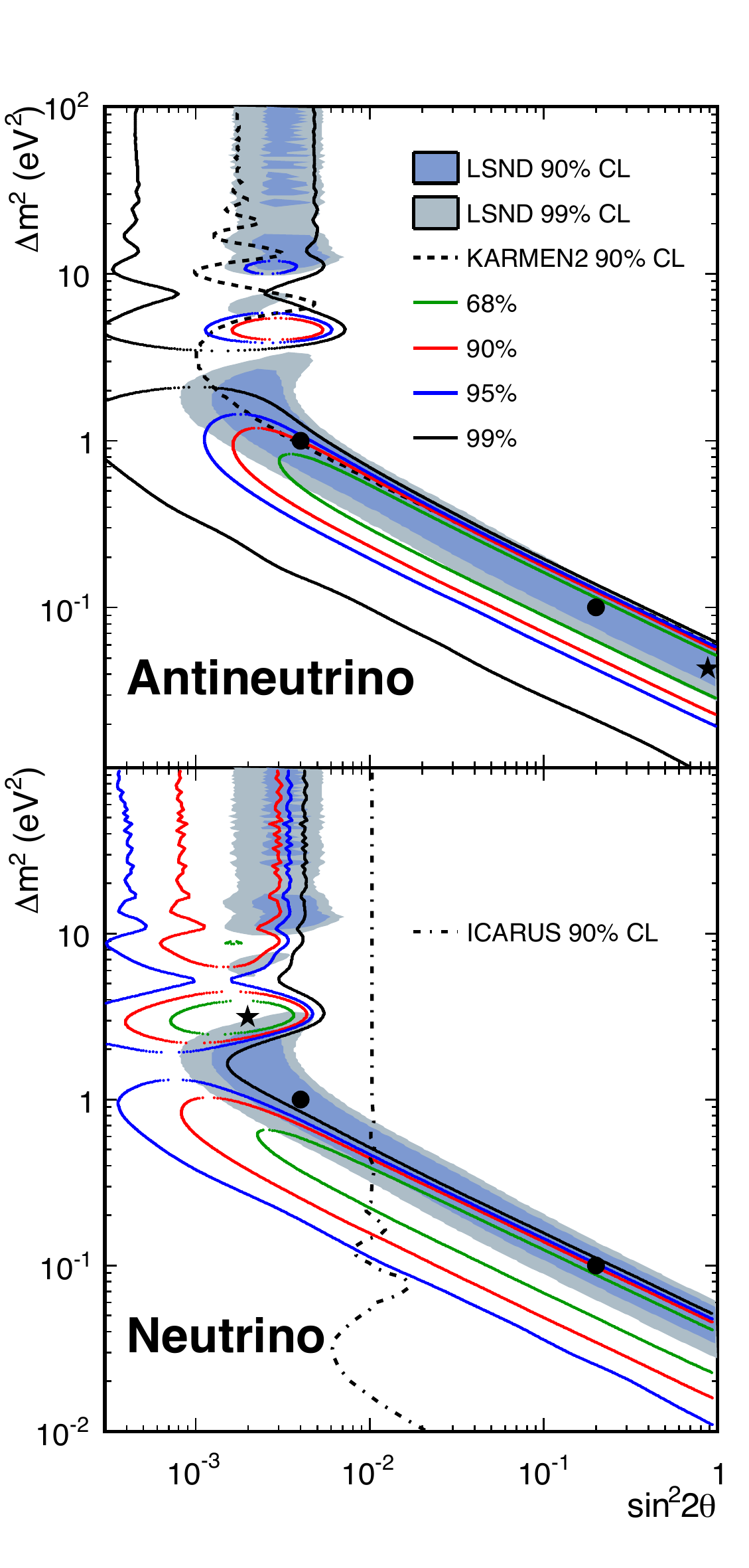}}
 \vspace{-0.3in}
\caption{MiniBooNE allowed regions in antineutrino mode (top) 
and neutrino mode (bottom) for events with
$E^{QE}_{\nu} > 200$ MeV within a two-neutrino oscillation model. 
Also shown are the ICARUS \cite{icarus} and KARMEN \cite{karmen} 
appearance limits for neutrinos and antineutrinos, respectively.
The shaded areas show the 90\% and 99\% C.L. LSND 
$\bar{\nu}_{\mu}\rightarrow\bar{\nu}_e$ allowed 
regions. The black stars show the MiniBooNE best fit points, 
while the circles
show the example values used in Fig.~\ref{excessnab}.}
\label{limitab}
\vspace{-0.1in}
\end{figure}

The MiniBooNE antineutrino data can be fit to a two-neutrino oscillation model, where the
probability, $P$, of $\bar \nu_\mu \rightarrow
\bar \nu_e$ oscillations is given by $P=\sin^22\theta \sin^2(1.27 \Delta m^2 
L/E_\nu)$, $\sin^22\theta = 4|U_{e4}|^2|U_{\mu4}|^2$, 
and $\Delta m^2= \Delta m^2_{41} = m^2_4-m^2_1$. 
The oscillation parameters are extracted from a combined fit of 
the observed $E_\nu^{QE}$ event distributions for muon-like and electron-like events.
The fit assumes the same oscillation probability for 
both the right-sign $\bar \nu_e$  and wrong-sign  $\nu_e$, and
no significant $\nu_\mu$, $\bar{\nu}_{\mu}$, $\nu_e$,
or $\bar \nu_e$ disappearance. 
Using a likelihood-ratio technique \cite{mb_osc_anti},  the confidence level
values for the fitting statistic, $\Delta\chi^2 =\chi^2(point)-\chi^2(best)$, 
as a function of oscillation parameters, $\Delta m^2$ and $\sin^22\theta$, is determined
from frequentist, fake data studies.  The critical values over the oscillation
parameter space are typically 2.0, the number of fit parameters, but can
be as a low as 1.0 at small $\sin^22\theta$ or large $\Delta m^2$.
With this technique, the best antineutrino oscillation fit 
for $200<E_\nu^{QE}<3000$~MeV occurs at
($\Delta m^2$, $\sin^22\theta$) $=$ (0.043 eV$^2$, 0.88)
but there is little change in probability in a broad region 
up to ($\Delta m^2$, $\sin^22\theta$) $=$ (0.8 eV$^2$, 0.004)
as shown in Fig. ~\ref{limitab} (top). 
In the neutrino oscillation energy range of
$200<E_\nu^{QE}<1250$~MeV, the $\chi^2/ndf$ for the 
above antineutrino-mode best-fit point
is 5.0/7.0 with a probability of 66\%.
The background-only fit has a $\chi^2$-probability of
0.5\% relative to the best oscillation fit and a $\chi^2/ndf = 16.6/8.9$
with a probability of 5.4\%.
Fig.~\ref{limitab} (top) shows the MiniBooNE closed 
confidence level (CL) contours for
$\nu_e$ and $\bar \nu_e$ appearance oscillations in 
antineutrino mode in the
$200<E_\nu^{QE}<3000$~MeV energy range. 
The data indicate an oscillation signal region at the greater than
99\% CL with respect to a no oscillation hypothesis, 
which is consistent with some parts of the LSND 99\% CL allowed
region and consistent with the limits from the KARMEN experiment \cite{karmen}.  

Multinucleon processes and $\nu_e$ and $\nu_\mu$ disappearance can affect the 
results of the MiniBooNE oscillation analysis.
Specifically, nuclear effects associated with neutrino interactions 
on carbon can affect the reconstruction of the
neutrino energy, $E_\nu^{QE}$, and the determination 
of the neutrino oscillation parameters \cite{Martini,Mosel,Nieves}.
These effects can change the visible energy in the detector 
and the relative energy distribution for the signal and gamma backgrounds. 
These effects are partially removed in this analysis 
since the gamma background is determined from
direct measurements of NC $\pi^0$ and dirt backgrounds. 

In order to estimate the possible effects
of a multinucleon-type model,
an oscillation fit was performed using event predictions based 
on the Martini {\it et al.} \cite{Martini} model.  The prediction was
implemented  by smearing the input neutrino energies as a function
of reconstructed energy to mimic the behavior of the model. 
For an estimate of the
effects of disappearance oscillations, a (3+1) type model
was used.  Fits were performed 
where the appearance $\Delta m^2$ and $\sin^2 2\theta_{app}$ 
parameters were 
varied as usual but disappearance oscillations were also included
with  $|U_{e4}|^2=|U_{\mu 4}|^2=|U|^2=\sqrt{\sin^2 2\theta_{app}/4}$
and with the same $\Delta m^2$ .  
This is a disappearance model where all four types of 
neutrinos ($\nu_e / \bar\nu_e / \nu_\mu / \bar\nu_\mu$)
disappear with the same effective $\sin^2 2\theta_{disapp} = 4(1-U^2)U^2$.  
A comparison of the results for these models versus the nominal 
MiniBooNE analysis is given in Table  \ref{model_studies}.  
Results are presented for the best fit with the given prediction model and
for a test point with $\Delta m^2 = 0.5$ eV$^2$ and $\sin^2 2\theta = 0.01$.  
The difference in $\chi^2$ values for the different prediction models
is $<0.5$ units, suggesting that multinucleon or disappearance 
effects do not significantly change the oscillation fit and null exclusion
probabilities.  

\begin{table}[t]
\vspace{-0.1in}
\caption{\label{model_studies} \em $\chi^2$ values from oscillation fits 
 to the antineutrino-mode data for different prediction models.   The best fit
 ($\Delta m^2, \sin^2 2\theta$) values are 
 (0.043 eV$^2$, 0.88), (0.059 eV$^2$, 0.64), and (0.177 eV$^2$, 0.070) for 
 the nominal, Martini, and disappearance models, respectively.  
 The test point $\chi^2$ values in the third column are for 
 $\Delta m^2 = 0.5$ eV$^2$ and $\sin^2 2\theta = 0.01$.  The effective
 $dof$ values are approximately 6.9 for best fits and 8.9 for the
 test points.
 }
\small
\begin{ruledtabular}
\begin{tabular}{l|cc}
   & \multicolumn{2} {c} {$\chi^2$ values}     \\
Prediction Model& Best Fit & Test Pt.  \\  \hline

Nominal $\bar\nu-$mode Result & 5.0 & 6.2   \\

Martini {\it et al.} \cite{Martini} Model&  5.5    &   6.5    \\

Model With Disapp. (see text) &  5.4    &   6.7    \\

\end{tabular}  
\vspace{-0.2in}
\end{ruledtabular}
\normalsize
\end{table}

Even though the MiniBooNE antineutrino data is a direct test of the 
LSND oscillation hypothesis, the MiniBooNE neutrino-mode
data can add additional information, especially for comparisons to various sterile neutrino
models.  The  previous MiniBooNE oscillation analysis \cite{mb_osc} found no evidence for neutrino oscillations in
neutrino mode by fitting over the neutrino energy range $475<E_\nu^{QE}<3000$~MeV, excluding the
low-energy range, $200<E_\nu^{QE}<475$~MeV.  The reason for excluding the low-energy region in the original analysis 
was based on uncertainties for the large gamma background in that region.  The subsequent work on constraining the low energy
background and making a more accurate assessment of the uncertainties now allow the data below 475 MeV to be used \cite{mb_lowe}.
The neutrino-mode event  and excess distributions for $6.46 \times 10^{20}$ POT
are shown in the bottom plots of Figs. \ref{excessnat} and \ref{excessnab}, respectively.
In neutrino mode, a total of 952 events are in the region with $200<E_\nu^{QE}<1250$~MeV, 
compared to a background expectation of $790.1 \pm 28.1(stat.) \pm 38.7(syst.)$ events. 
This corresponds to a neutrino-mode excess of $162.0 \pm 47.8$ events
with respect to expectation or a 3.4$\sigma$ excess. 

Two-neutrino oscillation model fits to the MiniBooNE neutrino-mode data 
do show indications of oscillations as shown in Fig.~\ref{limitab} (bottom).
In contrast to the antineutrino-mode results, the MiniBooNE favored neutrino-mode region 
has only small overlap with the LSND region and
may indicate that the compatibility between the two is low in a simple two-neutrino model.
The best neutrino oscillation fit occurs at 
($\Delta m^2$, $\sin^22\theta$) $=$ (3.14 eV$^2$, 0.002).
In the neutrino oscillation energy range of
$200<E_\nu^{QE}<1250$~MeV, the $\chi^2/ndf$ for the best-fit point
is 13.2/6.8 with a fairly small probability of 6.1\%, and the background-only fit
has a  $\chi^2$-probability of 2\%  relative to the best oscillation fit 
and a $\chi^2/ndf = 22.8/8.8$ with a probability of 0.5\%.
As shown in Fig.~\ref{excessnab} (bottom), the poor $\chi^2/ndf$ for the 
neutrino-mode best fit
is due to the data being higher than the expectation at low energy and lower at
high energy.  This may be due to the limitation of the simple two-neutrino 
model if the excess is due to oscillations or
to some anomalous background at low energy 
if the excess is related to backgrounds.

In summary, the MiniBooNE experiment 
observes a total event excess in antineutrino mode running 
of $78.4 \pm 28.5$ events (2.8$ \sigma$) in the 
energy range $200<E_\nu^{QE}<1250$~MeV.  
The allowed regions from a two-neutrino fit to the data, shown in Fig. \ref{limitab} (top), 
are consistent with $\bar{\nu}_{\mu}\rightarrow\bar{\nu}_e$ oscillations 
in the 0.01 to 1 eV$^2$ $\Delta m^2$ range 
and have some overlap with the allowed region reported by the LSND 
experiment \cite{lsnd}.   All of the major backgrounds are constrained by in-situ 
event measurements so non-oscillation explanations would need to invoke new anomalous
background processes.  The neutrino mode running also shows an excess
of $162.0 \pm 47.8$ events ($3.4 \sigma$), but the energy
distribution of the excess is marginally compatible with a simple two neutrino oscillation
formalism.  While this incompatibility might be explained by unexpected systematic 
uncertainties and backgrounds, expanded oscillation models with several 
sterile neutrinos can reduce the discrepancy by allowing for CP 
violating effects.  On the other hand, global fits  \cite{3+2} with these expanded models
show some incompatibility with the current upper limits on electron and
muon neutrino disappearance that will need new data and studies to resolve.

\begin{acknowledgments}
We acknowledge the support of Fermilab, the Department of Energy,
and the National Science Foundation, and
we acknowledge Los Alamos National Laboratory for LDRD funding. 
\vspace{-0.01in}
\end{acknowledgments}



\end{document}